\documentclass[12pt,preprint]{aastex}

\usepackage{graphics,color}
\usepackage{graphics}
\usepackage{epsfig}
\usepackage{amsmath}
\usepackage{amssymb}

\begin{document}
\newcommand {\CO} {COMPASS }
\newcommand {\COp} {COMPASS. }
\newcommand {\PO} {POLAR }

\title{\CO: An Upper Limit on CMB Polarization at an Angular Scale of
$20 \arcmin$} 

\author{Philip C. Farese\altaffilmark{1}, 
Giorgio Dall'Oglio\altaffilmark{2}, 
Joshua O. Gundersen\altaffilmark{3},
Brian G. Keating\altaffilmark{4}, 
Slade Klawikowski\altaffilmark{5}, 
Lloyd Knox\altaffilmark{6},
Alan Levy\altaffilmark{7},
Philip M. Lubin\altaffilmark{7}, 
Chris W. O'Dell\altaffilmark{8},
Alan Peel\altaffilmark{6}, 
Lucio Piccirillo\altaffilmark{9}, 
John Ruhl\altaffilmark{10}, 
Peter T. Timbie\altaffilmark{5}}

\begin{abstract}
\CO  is an on-axis 2.6 meter telescope coupled to a correlation
polarimeter operating at a wavelength of 1 cm.  The entire instrument
was built specifically for CMB polarization studies.  We report here on
observations of February 2001 - April 2001 using this system.  
We set an upper limit on E-mode polarized anisotropies of 33.5 $\mu\text{K}$ 
(95\% confidence limit) in the $\ell$-range 200-600. 
\end{abstract}



\altaffiltext{1}{Department of Physics, University of California, 
Santa Barbara, CA 93106; \newline Present Location: Princeton University}
\altaffiltext{2}{Univerity of Rome, Rome Italy}
\altaffiltext{3}{Department of Physics, University of Miami, Coral Gables, FL 33146}
\altaffiltext{4}{Division of Physics, Math, and Astronomy, California Institute of
  Technology, Pasadena, CA 91125}
\altaffiltext{5}{Department of Physics, University of Wisconsin, Madison, WI 53706} 
\altaffiltext{6}{Department of Physics, University of California,
  Davis, CA 95616}
\altaffiltext{7}{Department of Physics, University of California,
  Santa Barbara, CA 93106} 
\altaffiltext{8}{Christopher W. O'Dell Department of Atmospheric and
  Oceanic Sciences, \newline University of 
Wisconsin-Madison, Madison, WI, 53706}
\altaffiltext{9}{Department of Physics and Astronomy, University
of Wales - Cardiff, Wales, UK CF24 3YB}
\altaffiltext{10}{Department of Physics, University of California, 
Santa Barbara, CA 93106; \newline Present Location: Case Western University}

\section{Introduction}
The recent detection of acoustic peaks in the cosmic microwave
background (CMB) temperature power spectrum supports the scenario
that we live in a critical density universe consisting mainly of
dark matter and dark energy. In this model, large-scale structures
grew via gravitational instability from seeds laid down by quantum
fluctuations in a period of inflation in the early universe \citep{hu02}.
Concordance models, in which the anisotropy
measurements are combined with observations of the large-scale
structure in the universe and observations of distant supernovae,
sharpen this picture even further.

Further clues from the early universe have recently been found in
the polarization of the CMB. The polarization signal is typically
divided into two types: E-modes, which arise from scalar (density)
perturbations in the early universe and B-modes, which are caused
by tensor (gravitational wave) perturbations.  The first attempts
to measure CMB polarization occurred shortly after the discovery
of the CMB almost 40 years ago. With the detection of the acoustic
peaks in the temperature power spectrum now fully in hand, a
variety of experiments have focused on polarization measurements.
The DASI experiment has recently detected the E-mode signal at
angular scales of $1^{\circ} - 0.2 ^{\circ}$\citep{Kovac02} at a level
of $\sim 5 \mu\text{K}$. DASI also detected the temperature-polarization
(TE) cross-correlation signal. Both detections are consistent with
predictions from ${\rm \Lambda CDM}$ models. Because the small-scale
polarization signal is expected to arise 
from the same processes that produce the acoustic peaks in the
temperature anisotropy, detection of this signal has provided
reassuring confirmation of the temperature measurements and
further support for the inflationary scenario. WMAP has detected the
TE polarization signal on angular scales $ > 0.2 ^{\circ}$
\citep{Kogut03}. On scales below $5^{\circ}$ the signal is consistent
with that expected from the observed temperature power spectrum.
But on large scales $>10^{\circ}$, excess power is detected that
is consistent with reionization occurring in the redshift range
$11 <z_{r} <30$ with an optical depth of $\tau = 0.17 \pm 0.04$.
Further
measurements of the polarization power spectrum will improve the
determination of the fundamental cosmological parameters. In
particular, measurement of the B-mode signal will constrain or
detect primordial gravitational waves created during inflation.

We report here a 95 \% confidence upper limit at an angular scale
of $\sim 0.3^{\circ}$ in the frequency range 26-36 GHz from one
season of observations with the COMPASS (COsmic Microwave
Polarization at Small Scales) telescope.  Although this limit is
about a factor of 6 above the level of the recent polarization
detections at this scale, the measurements are important for
understanding the effects of foreground emission.  Foreground
emission from galactic and extragalactic sources can cause both E-
and B- mode contamination near the level of sensitivity we have
achieved. In our range of observing frequencies synchrotron
radiation and emission from spinning dust are expected to be the
dominant foregrounds.   COMPASS observes near the North Celestial Pole
(NCP).  While the 
NCP is an exceptionally convenient region of the sky to observe
from the Northern Hemisphere, it is not particularly clean of
potential galactic foreground radiation. The absence of any
detection of foreground signal in this region is encouraging for
future measurements.

This paper is organized as follows. In section 2 we describe the
COMPASS instrument, in section 3 the calibration, and in section 4
the observations. Sections 5, 6, 7, and 8 explain the data selection
and analysis procedures and section 9 gives the results.

\section{Instrument}

\CO uses a receiver that was originally coupled to a corrugated feed horn with a
$7 \arcdeg$ FWHM beam for a large angular scale CMB polarization
experiment known as \PO \citep{keating03}.  In order to
observe smaller angular scales where a larger primordial polarization
signal is expected a dielectric lens was added to the \PO optical
system and this horn+lens combination was coupled to a 2.6-meter
on-axis Cassegrain telescope to form \COp   Here we present an overview of the instrument; 
further instrumental details can be found in
\citep{farese03} and \citep{farese02}.

\subsection{Polarimeter}
The \CO polarimeter implements state-of-the-art HEMT (High-Electron Mobility
Transistor) amplifiers that operate in the 26-36 GHz frequency range (Ka band).  
These amplifiers are maintained at $\approx 20$ K and provide coherent
amplification with a gain of $\approx 25$ dB  
and a noise temperature of $\approx 20$ K.  To reject $1/f$ noise
from the detectors and atmospheric fluctuations two HEMTs are
configured as a correlation receiver with an AC phase modulation.  
Each HEMT amplifies one of two polarizations observed through the same horn and thus the
same column of atmosphere and same location on the sky at a given
time.  The resultant amplified signals are mixed down to 2-12 GHz,
phase-modulated at 1 kHz, multiplexed into three sub-bands, and
further amplified along separate but identical IF amplifier
chains. The signals are then multiplied together and the 1 kHz phase switch
is demodulated.  The resulting correlated time-averaged signal
amplitude is proportional to one linear combination of the Stokes
parameters (Q or U in the frame of the polarimeter) as determined by the
parallactic angle of the observations and the orientation of the
receiver axes.  A second linear combination can be obtained after a
$45 \arcdeg$ rotation of the polarimeter about the feedhorn's optical axis,
which allows one to measure both Stokes parameters and thus obtain
all information about the linear polarization.  For the observations
reported here no rotation was performed; solely the polarimeter's U
polarization was measured.  Depending on the observation strategy the
polarimeter signal will then map to some linear combination of Q and U
Stokes parameters, and similarly E and B modes, on the celestial sphere.

The three sub-bands, in order of decreasing RF frequency, are termed
J1, J2, and J3. The noise of each sub-band, including telescope
efficiency and atmospheric absorption (but not emission) effects, 
was 1040, 850, and 820 ($\mu$K$\sqrt{\text{s}}$) respectively.
Each of these sub-bands is demodulated with waveforms
that are both in-phase and out-of-phase with the phase-modulation signal.  The
desired signal is obtained from each in-phase demodulation (called
J1I, J2I, and J3I) in addition to a null-signal noise monitor from the
out-of-phase demodulation (J1O, J2O, and J3O).  Further, a power
splitter prior to the multiplexing stage allows a total power
detection of each linear polarization termed TP0 and TP1;  
this signal is used for diagnostic purposes only.  
Additional details regarding the polarimeter can be found in 
\citep{keating03}.  

\subsection{Telescope}
The \CO optics were designed to be as free from systematic effects as
possible.  Oblique reflections of light off metallic surfaces induce
spurious polarization. Scattering or diffraction by a metal or
dielectric will also induce a polarized signal \citep{kildal88}.  In
any off-axis telescope there will be a systematic polarization for at
least one Stokes 
parameter.  In conventional on-axis systems metallic or dielectric
secondary supports necessarily obstruct the optical aperture and
give rise not only to an overall polarization but also an increased
side-lobe level.  To avoid both of these effects \CO uses a
microwave-transparent expanded polystyrene (EPS) secondary support system.
This support was designed to position and stabilize the secondary
mirror to 1 mm accuracy \citep{farese03}.

The secondary mirror was designed to minimize aperture blockage.  The
2.6 m diameter primary mirror had a 30 cm diameter hole in its
center, so the secondary was constructed with a 30 cm diameter.  A
hole was left in the center of the secondary to prevent
re-illumination of the receiver.  A polarized calibrator made of a
thermal source and wire grid was placed behind this hole.  The
secondary mirror is designed to under-fill the primary mirror; the
primary illumination edge illumination ends 7 cm from the edge of the
primary with a much faster than Gaussian taper.  This results in
reducing ground pickup and spillover while still maintaining a small
beam size.


In order to minimize the illumination at the edge of the secondary
mirror using the existing microwave horn antenna and dewar it was necessary to
reduce the beam size from $7 \arcdeg$ FWHM to $5 \arcdeg$ FWHM with a
lens. The additional requirement that this lens be cooled to reduce
its contribution to the system temperature necessitated that the lens
be mounted close to the horn and thus be approximately the same
diameter as the horn.  A simple meniscus phase-correcting lens was
selected which resulted in -15 dB secondary edge illumination.  This
design was based on the work by \citep{kildal88} and has a spherical
inner surface, whose radius matches that of the radius of the
horn-produced Gaussian beam, and an ellipsoidal outer surface designed
to give a flat phase front at the entrance surface of the lens.

To further reduce the sensitivity of \CO to possible systematic effects two
levels of ground screens, one affixed to the telescope and the other
stationary, were constructed.  The screens  mounted to the
telescope were attached directly to the edge of the primary mirror.
These provide an additional $> 50$ dB of attenuation to signals from
the ground, Sun, and Moon in addition to the low sidelobe
level (-60 dB) of the telescope at the location of these contaminants.
Observations were conducted both with and without the outer
(stationary) ground screens present.  Data collected with the outer
groundscreens present suffered from a larger scan synchronous signal (SSS)
than data taken with them absent.  It is believed that the combination
of telescope spillover with the oblique scattering angle off the
metallic surface of the groundscreens induced a large polarized offset.
This will be discussed in more detail in section \ref{text:SSS}.

\CO uses a standard Az-El pointing platform.  The Azimuth and
Elevation stages are separate units.  The Azimuth table used was a
refurbished and improved version of an existing table previously
designed for use with CMB observations from the South Pole.  The
Elevation stage was designed and built for this experiment.  The
position of each axis was read out by a 16 bit encoder giving a
resolution of $20\arcsec$.  Data acquisition of all radiometer,
pointing, and ``house keeping'' data (such as thermometry) as well as
telescope control were performed with a Pentium class laptop computer
and 48 channel 16-bit ADC.

\subsection{Observing Site}
Our observations took place at the University of Wisconsin's Astronomy
Observatory in Pine Bluff, Wisconsin (89.685$\arcdeg$ West longitude,
43.078$\arcdeg$ north latitude). The telescope was housed in a 20 m
$\times$ 15 m tensioned fabric building with a wheeled, aluminum
frame.  This building was rolled 20 m to the South of the telescope on
tracks for observations and rolled over the telescope for shelter
during periods of foul weather.  By moving the building rather than
the telescope we were assured of the stability of the telescope and
its celestial alignment.


\subsection{Pointing}
We initially pointed the telescope by co-aligning an optical telescope
mounted on the primary mirror with the cm-wave beam pattern using a 31
GHz Gunn oscillator mounted on a radio tower 1.9 km West-by-Southwest
from the telescope.  The source on the radio tower was quite easily
visible through this optical telescope.  Because of the proximity of our
observing region to the NCP several optical observations of Polaris
and a number of nearby stars were made.  These Polaris observations
were then used to define our absolute Azimuth and Elevation offset and
thus to define our observing region.  Unfortunately, misalignment of
the optical telescope resulted in an Azimuth and Elevation pointing
error of $+0.458\arcdeg \pm 0.027\arcdeg$ and 
$-0.107\arcdeg \pm 0.023\arcdeg$ respectively.

This pointing offset was discovered through observations of the
supernova remnant Cas A.  Cas A is circumpolar from our observing
location, fairly close to our observing region, and quite bright in
intensity, making it useful as a pointing tool.  Three observations
were made throughout the season as outlined in Table \ref{tab:casA}.
Through linear fits the Elevation is seen to drift at $-1.43\arcmin
\pm 2.3\arcmin$ in Elevation and $10.1\arcmin \pm 6.9\arcmin$ in
Azimuth per month; There is no compelling statistically significant
evidence for drifts in the pointing over the given 52 day period.
Thus, a fixed pointing offset was used for all files in the analysis.

\begin{table}[!t]
\begin{center}
\begin{tabular}{|c|c|c|c|c|c|} \hline
Day & Obs. & Az   & error &  El   & error \\ \hline
62  & 4    & 17.4$\arcmin$ & 6.6$\arcmin$   & -5.4$\arcmin$  & 4.2$\arcmin$ \\ \hline
109 & 2    & 40.8$\arcmin$ & 1.8$\arcmin$   & -10.2$\arcmin$ & 2.4$\arcmin$ \\ \hline
114 & 6    & 29.4$\arcmin$ & 4.3$\arcmin$   & -6.0$\arcmin$  & 1.8$\arcmin$ \\ \hline \hline
Ave &      & 37.8$\arcmin$ & 1.6$\arcmin$   & -7.3$\arcmin$  & 1.4$\arcmin$ \\ \hline
Uncert. &  & 5.3$\arcmin$  &       &  1.8$\arcmin$  & \\ \hline
\end{tabular}
\caption[Cassiopeia A: Pointing]{Successful observations of
Cas A were made on the days specified by the day number in the
first column.  ``Obs.'' gives the number of full, independent rasters
performed at that time.  The Azimuth and Elevation offsets and errors
are given in arcminutes in the remaining columns.  ``Ave'' provides the
weighted average, and thus the offsets used in analysis and
``Uncert.'' the scatter weighted by the uncertainty of each
measurement.} 
\label{tab:casA}
\end{center}
\end{table}

\subsection{Beam Determination}
Raster maps of the above-mentioned ``tower source'' were made by using
an iso-Elevation raster scan.  This source was essentially in the far
field of the telescope, requiring only a 0.1$\arcmin$ correction to
the beamwidth at infinity.  The FWHM obtained by fits of a Gaussian to
the Elevation and Azimuth directions for two separate days of tower
source observations are 19.2$\arcmin$ $\pm$ 0.4$\arcmin$ and
20.3$\arcmin$ $\pm$ 0.5$\arcmin$ respectively.  These beam maps are
confirmed by scans of Tau A.  A fit of these maps with Tau A
deconvolved yield a FWHM for each polarization channel and each
direction (Elevation and cross-Elevation) as provided in Table
\ref{tab:FWHM_Tau}.  Note that the analysis of the Tau A data requires
time series filtering which may induce a larger beam size. These
numbers are further confirmed by observations of Venus which is a
point source in our beam and yields beams of the two total power
channels of $18.5^{'} \pm 1.0^{'}$ and $19.6^{'} \pm 0.8^{'}$.  In our
likelihood analysis we make the approximation that the beam is axially
symmetric with FWHM $20.0^{'}$, an approximation which affects our
result by $\le$ 5\%.

\begin{table}[!ht]
  \begin{center}
    \begin{tabular}{|c|c|c|c|} \hline
      Method & Channel     & Cross-Az & Elevation \\ \hline
      Tau A  & J1I (32-35) & $21.5\arcmin \pm 0.9\arcmin$ & 
      $20.2\arcmin \pm 0.9\arcmin$ \\ \hline
      Tau A  & J2I (29-32) & $22.0\arcmin \pm 0.7\arcmin$ & 
      $21.5\arcmin \pm 0.9\arcmin$ \\ \hline
      Tau A  & J3I (25-29) & $23.2\arcmin \pm 0.7\arcmin$ & 
      $23.2\arcmin \pm 0.9\arcmin$ \\ \hline
      Tower  & TP0, TP1    & $19.2\arcmin \pm 0.4\arcmin$ & 
      $20.3\arcmin \pm 0.5\arcmin$ \\ \hline
      Venus  & TP0, TP1    & $18.5\arcmin \pm 1.0\arcmin$ & 
      $19.6\arcmin \pm 0.8\arcmin$ \\ \hline
    \end{tabular}
  \end{center}
  \caption[Beam Widths]{FWHM of the beam on the sky of each sub-band 
    derived from observations of Tau A, Venus, and our ``tower source''.}
  \label{tab:FWHM_Tau}
\end{table}

\section{Calibration}
Given our $\sim20\arcmin$ beam and our sensitivity level, we
require a source that is no more extended than several arcminutes and
of $\sim10-20$ Jy polarized power when averaged 
over our telescope's beam. Tau A (the Crab Nebula) meets these
requirements and is our primary calibration source.

Historically there have been many observations of the Crab Nebula at
many frequencies.  If all are taken at face value they are not 
in agreement with each other.  To narrow the field we have chosen
to use only observations that were within our observing band. These
observations are corrected by the spectral index of Baars
($\alpha=-0.299 \pm 0.009$ \citep{baars77} and the observed decay rate
($-0.167\% \pm 0.015 \%/$year) of Allers \citep{aller85} to
contemporary, 30 GHz observations.  Seven measurements satisfy these
criteria (see Table \ref{tab:cali}), resulting in a most likely total
power flux measurement of $339 \pm 11$ Jy.  These figures
remained quite robust to reasonable variations in the subset of data
selected and assumptions about the reported data and errorbars.

Next a total of eight polarized measurements are combined to obtain a
most likely estimate of the polarized fraction of Tau A as $8.1 \pm
0.9$ \%.  For this analysis frequency has been ignored for there is no
observed frequency dependence of the data.

Finally the parallactic angle, beam dilution, and atmospheric
absorption of the source at the time of each observation as well as the
frequency dependence of the three correlator channels must be taken
into account.  A Gaussian beam of $20\arcmin$ (38.4
$\mu\text{radians}^2$ was used.  The source is observed by performing
an iso-Elevation 
raster scan over the target region.  A linear term is fit and removed from each
raster scan and the residual scans combined and compared to a template
of the source derived from \citep{Johnston69} to obtain our
calibration.  All of these procedures are summarized in Table
\ref{tab:cali}. 

\begin{table}[!t]
\begin{center}
\begin{tabular}{|c|c|c|c|c|c|c|c|c|c|} \hline
Meas. & mFlux & Err.& Year & Decay & Freq.& Cor. &
cFlux & Pol. \% & Par.\\ \hline
\tiny \citep{Effels}    & 304 &	31 &	'98 &	0.995 &	32   &	1.02 &
308 &	8.3 & 152 \\ \hline
\tiny \citep{Johnston69}& 313 &	50 &	'69 &	0.947 &	31.4 &	1.01 &
300 &	8.1 & 158\\ \hline
\tiny \citep{Hobbs68}  &	387 &	72 &	'68 &	0.945 &	31.4 &	1.01 &
371 &	xx & xx\\ \hline
\tiny \citep{Kalaghan67}  &	340 &	53 &	'67 &	0.943 &	34.9 &	1.05 &
336 &	12 & 140 \\ \hline
\tiny \citep{CBI}   & 355 &   xx &	'98 &	0.995 &	31   &	1.01 &	357 &
7 & 152 \\ \hline
\tiny \citep{Allen67}& 373 &	xx &	'69 &	0.947 &	31.4 &	1.01 &	358 &
xx & xx\\ \hline
\tiny \citep{Hobbs69} & 357 &	xx &	'69 &	0.947 &	31.4 &	1.01 &	343 &
xx & xx\\ \hline
\tiny \citep{Mayer68}  &  xx &	xx &	xx &	xx &	19   &	xx &
xx &	6.6 & 152\\ \hline
\tiny \citep{Green75} &  xx &	xx &	xx &	xx &	15   &	xx &	xx &
6.4& 148\\ \hline
\tiny \citep{Boland66}   &  xx &	xx &	xx &	xx &	14.5 &	xx &
xx &	8.6 &146\\ \hline
Ave. & \multicolumn{3}{|c|}{Flux (Jy): 339 $\pm$ 11} & 
\multicolumn{3}{|c|}{Pol. \%: 8.9 $\pm$ 0.9} & 
\multicolumn{3}{|c|}{Par.: 150 $\pm$ 2} \\ \hline
\hline
Chan &	Freq &	Cor. &	RJ-T &	\multicolumn{2}{|c|}{mK (sig)} & 
\multicolumn{2}{|c|}{mV (sig)} & \multicolumn{2}{|c|}{Gain (K/V)} \\ \hline
J1 &	34   &	0.96 &	1.02 &	\multicolumn{2}{|c|}{11.9 $\pm$ 3.5} & 
\multicolumn{2}{|c|}{3.6  $\pm$ 0.16} & \multicolumn{2}{|c|}{3.32
  $\pm$ 0.98}\\ \hline
J2 &	30.5 &	1.00 &	1.02 & \multicolumn{2}{|c|}{15.3 $\pm$ 3.2} & 
\multicolumn{2}{|c|}{5.2 $\pm$ 0.14} &\multicolumn{2}{|c|}{2.98 $\pm$ 0.62} \\ \hline
J3 &	27.5 &	1.03 &	1.03 & \multicolumn{2}{|c|}{19.6 $\pm$ 4.0} & 
\multicolumn{2}{|c|}{12.4 $\pm$ 0.37} &\multicolumn{2}{|c|}{1.58 $\pm$ 0.33}\\ \hline
\end{tabular}
\caption[Calibration]{Previous measurements of Tau A near 30 GHz. In
  the top porting of the table ``Meas.'' indicates the reference for
  the measurement and ``Ave.'' indicates the most likely value given
  all information in this table. ``mFlux'' and ``Err.''  as measured by
  the indicated group are given in Jy.  ``Year'' indicates the year of
  the observation and ``decay'' indicates the correction for temporal
  signal decay between the indicated observed date and our own as
  calculated through \citep{aller85}. ``Freq.'' and ``Cor.'' are
  the frequency of the original observation and the correction to \CO
  central frequency as per \citep{baars77}.  ``cFlux'' indicates
  the flux that results from applying these two corrections.  ``Pol. \%'' and
  ``Par.'' are the polarized flux percentage and parallactic
  angle in degrees.

  In the second portion corrections for the individual channels are
  applied.  ``RJ-T'' is the Rayleigh-Jeans to Thermodynamic units
  correction, ``Omega'' is the solid angle of the beam in radians$^2$,
  ``mK'' is the expected signal in (thermodynamic) temperature units
  given  all the above and ``mV'' is the most likely amplitude of our
  observations.} 
\label{tab:cali}
\end{center}
\end{table}

\section{Observations}
\label{text:obs_strat}
Our observing strategy was designed to optimize the probability of
detecting a signal under current well-motivated theories while still
allowing for systematics checks and tests.  We attempted to observe a
circular ``disk'' region centered on the NCP by
maintaining a constant Elevation and scanning the telescope in
Azimuth.  During constant-Elevation scans the thermal load from the
atmosphere remains constant and reduces both gain fluctuations of the
receiver and intensity-polarization coupling in the polarimeter that
could cause systematic effects.  Our full scan period was varied
between 10 and 20 seconds to reject longer term atmospheric
fluctuations while maintaining stable and reliable telescope
performance.

\begin{figure}[!t]
\begin{center}
\includegraphics{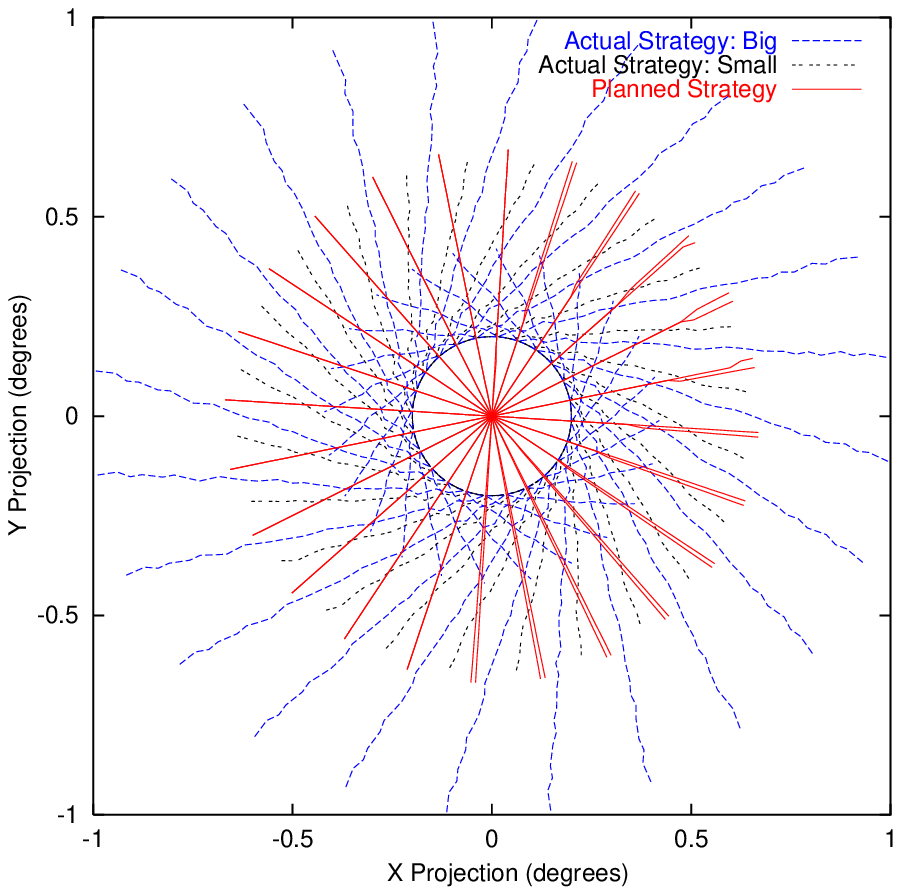}
\caption[Scan Strategy]{\CO scan strategy in rectangular projection
  about the NCP.  The axes are real degrees on the sky with the positive x-axis
  corresponding to a Right Ascension of 0h.  It is clear
  how constant Elevation Azimuth scans are transformed into a two
  dimensional map through sky rotation.} 
\label{fig:ScanStrat}
\end{center}
\end{figure}

As the sky rotates this scanned line is transformed into a cap
centered on the NCP as demonstrated in Figure \ref{fig:ScanStrat}.
Further, because of the sky rotation, each half sidereal day the same
region of sky is observed and allows the use of difference maps as a
robust test of systematic errors.  Initially this ``cap'' was one
degree in diameter in order to allow deep integration on a small patch
of sky to search for systematic effects.  Half way through the season
this diameter was increased to $1.8\arcdeg$ to reduce sample variance.  As
mentioned above 
there was a small pointing offset so our actual scan strategy was not
the one we intended.  This resulted in a loss of symmetry and thus
some systematic tests but acceptable noise properties.  Given the
relative sizes of our beam, scan region, and pointing offset from the
NCP our telescope is sensitive to both E and B modes
at roughly the same level. \label{text:scan_strat}

One further benefit of this scan strategy is that it simplifies the
process of encoding the time stream data into a map which displays
well the scan synchronous signal  (SSS) while still allowing for
reasonable cross-linking, sampling 
of a variety of parallactic angles, and systematic tests.
A one-dimensional to two-dimensional map comparison allows simple
identification of the SSS.  If we plot the data in a coordinate system
defined as Azimuth and Right Ascension it becomes quite easy to
combine the data across many days and project out the modes that
correspond to the SSS (see section \ref{text:SSS}).  

Finally, two configurations of the ground screens were implemented; one
utilizing the two layers mentioned above and one with only the inner,
co-moving ground screen.  These two options, combined with the two
scan amplitudes mentioned above, result in a total of four ``sub-seasons" of
data termed SIGS, SOGS, BIGS and BOGS to specify whether the (S)mall
or (B)ig scan was used and whether the (O)uter or only (I)nner ground
screens were in place.  Analysis was performed on each channel in
each sub-season as well as combinations of channels within a
sub-season, combinations of sub-seasons for a given channel, and all
acceptable data.

\section{Data Selection, Cleaning, and Reduction}
The data analysis pipeline consists of five steps: data selection,
data cleaning, intermediate Azimuth map making, full map making, and
power spectrum estimation.  The first three steps are outlined in this
section with the last two described in greater detail in the following
sections.  

There were a total of 1776 hours available during our observing season
which was defined as the months of March, April, and half of
May of 2001. Of these, 409 had sufficiently good weather to operate the
experiment.  74 hours were ignored because heavy winds disrupted
the Azimuth pointing and control and 28 hours were removed because of
equipment failures.  This leaves a total of 309 hours of usable data
observing the target region.  Further cuts are made, based on the data, to select periods
of stable observing conditions.

The data are divided into files of 15 minutes in length.  For each
file six statistics are generated: three 
noise-based statistics (white noise level, 1/f knee, and $\zeta$
\citep{keating03} -- essentially a $\frac{1}{f}$ weighted sum of the
PSD); two cross correlations (between two correlator channels
and between a correlator and total power channel); and a linear drift of
the time stream.  The 1/f knee and $\zeta$ statistics proved to be
most sensitive to periods of light cloudiness or haziness.  The cross
correlations were most useful in identifying more rapid contaminating
events such as discrete clouds, birds, or planes interfering with the
observations.  Finally the linear drift was used to identify dew
formation on the foam cone.

A histogram of each statistic is formed and unions and intersections
of files passing sets of cuts at the 3-$\sigma$ level are made.  Our
results were not overly sensitive to which sets were selected.  The
results reported here use the two cross-correlation criteria for they
retained the most data while still passing all null-tests performed
(see below). After all selection procedures the number of hours kept
were 144, 123, and 164 for J1, J2, and J3 respectively.

Each file is then passed through a despiking procedure.  Regions
with excessive slopes, second derivatives, or values greater than 5
standard deviations from the mean are flagged, cut, and filled with
white noise as estimated by the remainder of that file.  Such flagged
and white-noise filled data is not included in the data analysis. This
procedure removes between 0.1 \% and 32\% of the data from any given
sub-season. The 32\% was an anomaly; it occurred only for one channel,
J3, during the first sub-season where a damaged preamplifier gave rise
to excessive spikes in the data intermittently.  This was fixed once
the cause was identified.

As a first step in the analysis, we plot the data in each fifteen minute file 
as a function of Azimuth. The sky rotation on this time scale is sufficiently
small that a single bin of RA is needed for each file. This allows a great reduction
in data size as well as a simple treatment of our most likely
systematic effect (see section \ref{text:SSS}). In order to estimate
the noise in these files a single data point is formed for each
Azimuth bin of each Azimuth pass (i.e. leftward or rightward motion between two turn
around points in the telescope motion).  The standard error of the $\approx 100$
passes per file in each bin allow for an estimation of the noise to 10\% and was
in excellent agreement ($<$ 1\% different) with the noise used in a
full covariance estimation method.

\section{Map Making and Scan Synchronous Signal Effects}
\label{text:SSS}
One common systematic error in scanning style experiments is the
presence of non-celestial signal that correlates highly with position in the scan,
termed here ``Scan Synchronous Signal'' (SSS).  In \CO such a SSS was
observed and was related to a polarized offset induced by oblique
reflection from the stationary ground screen or spillover to the
ground.  We believe that this is the result of the aggressive
illumination of the secondary mirror.  In all sub-seasons a variable
offset and linear term (when plotted against Azimuth) are observed and
removed.  In the larger scan with the stationary ground screen present
a quadratic signal is detected and removed as well.  Once these
removals are performed the residuals are consistent with Gaussian
noise. 

Detection of the SSS is performed easily by comparing maps made by
binning data in only one dimension (i.e. Azimuth) to those made by
binning in two dimensions (i.e. Azimuth and Right Ascension).  If one has SSS
contamination the relationship between the reduced chi-squared of the
one dimensional map to that of the two dimensional map is given by
\begin{equation}
\chi^2_{1D}=1+N_{\text{bin}} \times (\chi^2_{2D}-1),
\end{equation}
where $N_{\text{bin}}$ is the number of bins in the dimension that is collapsed
and $\chi$ refers to reduced chi-squared.
As shown in Table \ref{tab:1D2D} there is no evidence for SSS
remaining in any of the sub seasons after removal of a first order
polynomial other than the Big Outer Groundscreen (BOGS)
configuration.  

\begin{table}[!ht]
\begin{center}
\begin{tabular}{|c|c|c|c|c|c|} \hline
& & \multicolumn{2}{|c|}{Two-D maps} 
& \multicolumn{2}{|c|}{One-D maps} \\ \hline
Season&Channel&DOF's& $\chi^2$&DOF's& $\chi^2$ \\ \hline 
SIGS &  J1I & 244 & 1.00 & 13 & 0.76 \\ \hline
SIGS &  J2I & 189 & 0.94 & 14 & 0.84 \\ \hline
SIGS &  J3I & 132 & 1.07 & 13 & 0.73 \\ \hline
SIGS &  J1O & 244 & 0.95 & 13 & 0.74 \\ \hline
SIGS &  J2O & 189 & 0.95 & 14 & 0.74 \\ \hline
SIGS &  J3O & 246 & 1.09 & 14 & 1.25 \\ \hline
SOGS &  J1I & 232 & 1.09 & 13 & 0.45 \\ \hline
SOGS &  J2I & 232 & 0.88 & 13 & 1.65 \\ \hline
SOGS &  J3I & 240 & 1.00 & 13 & 1.03 \\ \hline
SOGS &  J1O & 232 & 0.92 & 13 & 0.45 \\ \hline
SOGS &  J2O & 232 & 1.07 & 13 & 0.84 \\ \hline
SOGS &  J3O & 233 & 0.95 & 13 & 0.69 \\ \hline
BIGS &  J1I & 182 & 1.08 & 18 & 0.93 \\ \hline
BIGS &  J2I & 253 & 0.97 & 19 & 1.40 \\ \hline
BIGS &  J3I & 255 & 0.93 & 19 & 1.20 \\ \hline
BIGS &  J1O & 186 & 0.95 & 18 & 1.11 \\ \hline
BIGS &  J2O & 180 & 1.24 & 19 & 1.38 \\ \hline
BIGS &  J3O & 262 & 0.96 & 19 & 1.10 \\ \hline \hline
BOGS &  J1I & 217 & 1.24 & 19 & 3.96 \\ \hline
BOGS &  J2I & 216 & 1.22 & 19 & 4.52 \\ \hline
BOGS &  J3I & 212 & 1.17 & 19 & 4.20 \\ \hline
BOGS &  J1O & 217 & 1.10 & 19 & 1.44 \\ \hline
BOGS &  J2O & 216 & 1.01 & 19 & 0.68 \\ \hline
BOGS &  J3O & 217 & 1.03 & 19 & 0.74 \\ \hline
\end{tabular}
\caption[Map $\chi^2$ Tests]{The reduced Chi-squared tests for Az and RA-Az
binned maps for all subseasons and channels.  Note that all are
consistent with no signal except BOGS J1I, J2I, and J3I which all show
clear signs of unremoved Scan Synchronous Signal.}
\label{tab:1D2D}
\end{center}
\end{table}

\section{Map Making}
As a step toward estimating the CMB polarization power spectrum we
produce a map from our data. By the term ``map'' here we mean a
pixelized representation of the data which contains information on the
spatial location, most likely data value and (non-diagonal) noise
correlation matrix.  The nature of polarized observations
requires different mapping procedures than a simple intensity map.
Traditionally either a map of polarized intensity and orientation or a
map of the Q and U Stokes parameters is provided.  As \CO observed at
fixed Elevation the polarization direction information
(i.e. parallactic angle) is unambiguously encoded in the Right
Ascension and Azimuth.  

\begin{figure}[!t]
\begin{center}
\includegraphics{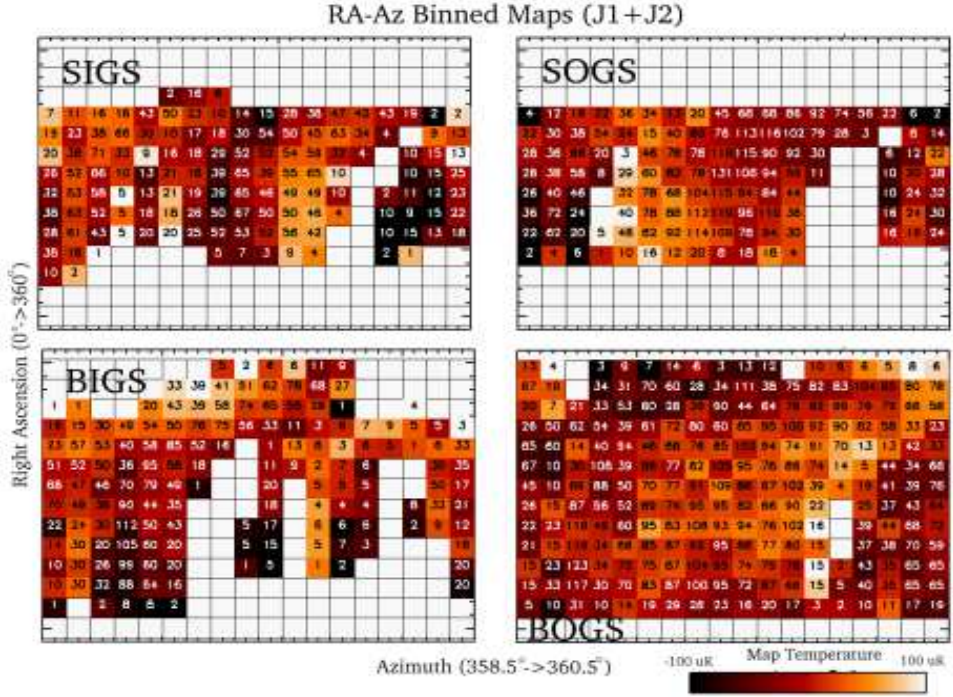}
\caption[RA-Az Maps]{Maps plotted in Right Ascension and Azimuth
coordinates.  Shown are maps for all 4 sub-seasons of the data for J1I
and J2I coadded. Color indicates intensity
of the signal per pixel while the number written over that pixel is
the number of Az-binned files that contribute to the value.}
\label{fig:maps}
\end{center}
\end{figure}

We map initially each (fifteen minute long) data file in Azimuth as
described above.  One useful map format in which to display the data
is RA-Az coordinates. This format facilitates identification of systematic
effects and spurious (or real) signals. Maps of the this style are
provided in Figure \ref{fig:maps}.  A second map format is a three
dimensional map of RA, Dec, and parallactic angle. This second map,
though sparsely populated, retains the polarization information of our
observations while still providing uniformly sized pixels and is used
in power spectrum estimation, though it is less intuitive to display
and interpret.

The noise covariance matrix is estimated from the timestream data.
The off-diagonal elements of this matrix were shown to be much less
than $10^{-6}$ of the amplitude of the on-diagonal elements and are
therefore ignored.  This is because the scan rate of $\approx 0.1$ Hz
is much slower than the anti-aliasing filter knee of 40 Hz.

Given this noise matrix and the knowledge that certain modes of the
map have been filtered through our polynomial subtraction method we
next produce a generalized noise covariance matrix by adding to the
noise matrix a constraint matrix as per \citep{bond98}.  The constraint
matrix encodes the SSS removal process and projects the removed modes
out of the map by adding noise of amplitude $10^8$ times that of the
noisiest on-diagonal element (units of $\mu K^2$) to the contaminated
modes.  These contaminated modes are constructed for each file's
Az-map and for each order of the fit removed by forming a constraint
vector and taking its outer product.  These constraint templates are
then multiplied by the indicated noise and added to the noise
covariance matrix of each file Az-map.  These sub-maps are then
added together in RA-Dec-parallactic angle space and a resultant
generalized covariance matrix, $C_N$, and map of the data, $D$, are formed
as given by
\begin{eqnarray}
C_{N,i}  &=& C_{n,i}+C_{c,i}, \nonumber \\
C_N^{-1} &=& \sum_i^M C_{N,i}^{-1}, \\
\vec{D} &=& C_N \left[ \sum_i^M \vec{d_i} C_{N,i}^{-1}. \right]
\nonumber
\end{eqnarray}
Subtraction of a constant term from each Azimuth file
resulted in weakening of our result (i.e. increasing the uncertainty
of the likelihood or 2-$\sigma$ limit) by 25\%.  Removing a constant term
and a slope weakened it by 50\%.

\section{Power Spectrum Estimation}
Software used to estimate the power spectrum was extensively tested
using simulated maps generated from a known power spectrum.  These
maps were provided by the UC Davis group and used to simulate a
\CO timeseries by the UCSB group.  Several hundred realizations of
this data set were produced and passed through the analysis
pipeline.  Both the most likely value and error estimation were
proved accurate and reliable for a wide range of constraint and
parallactic angle situations as shown in Table \ref{tab:pipe_check}.

\begin{table}[!ht]
\begin{center}
\begin{tabular}{|c|c|c|c|c|c|c|c|} \hline
& \multicolumn{5}{|c|}{Likelihood Probability} &
\multicolumn{2}{|c|}{\% Correct} \\ \hline
Test & $-2\sigma$ & $-1\sigma$ & Most Likely &$+1\sigma$ & 
$+2\sigma$ & \% w/i $1\sigma$ & \% w/i $2\sigma$ \\ \hline
Q only  & 17.6  & 18.8  & 20.0  & 21.5  & 22.9  & 56 & 92 \\ \hline 
U only  & 16.4  & 17.6  & 19.0  & 20.6  & 22.1  & 68 & 89 \\ \hline 
Q+U     & 14.5  & 16.3  & 18.1  & 20.8  & 23.3  & 63 & 92 \\ \hline 
Q+10    & 19.6  & 20.3  & 21.1  & 22.0  & 22.8  & 74 & 94 \\ \hline 
Q+20    & 17.5  & 18.0  & 18.5  & 19.0  & 29.5  & 74 & 99 \\ \hline 
con0    & 16.8  & 19.3  & 21.9  & 25.8  & 29.5  & 71 & 86 \\ \hline 
con1    & 14.4  & 16.3  & 18.3  & 21.1  & 23.9  & 63 & 93 \\ \hline 
con2    & 16.8  & 19.3  & 21.9  & 25.8  & 29.5  & 71 & 86 \\ \hline 
recov   & 15.1  & 17.1  & 19.3  & 22.4  & 25.5  & 72 & 95 \\ \hline 
\end{tabular}
\caption[Likelihood Tests]{Various test procedures applied to the
likelihood estimator and full pipeline.  The tests are described in
detail in the text. The first 5 use various combinations of Q and U
directly from the simulated maps.  The ``con'' tests create sets of
files of simulated Az-RA maps (our intermediate mapping) and combine
them optionally adding constraints (0=none, 1=DC removed per file,
2=DC and slope removed per file). recov uses a simulated time stream
generated from the simulated maps to test the full pipeline.}
\label{tab:pipe_check}
\end{center}
\end{table}
A flat band power model of E-mode polarization  was used to generate
the theory covariance matrix, $C_T(E)$, 
though use of a concordance model polarization spectrum did not
significantly change our results.  The likelihood of the amplitude of
this flat band power is calculated using the signal to noise eigenmode
method as described in \citep{bond98}.  It is worth noting that no
eigenmode in our data set has an eigenvalue greater than 1.  We build
the total covariance matrix $C(E)=C_T(E)+C_N$ and compute the
likelihood as given by  
\begin{equation}
L(\vec{D} | C(E))=\frac{1}{(2 \pi)^{N/2}} 
        \frac{e^{-\frac{1}{2}\vec{d}^T C(E)^{-1} \vec{d}}}
        {\left| C(E) \right|^{1/2}}.
\end{equation}
This likelihood is calculated on a grid spaced uniformly in units of
$\mu K^{2}$ to avoid over-biasing larger powers and is allowed to
take values of negative power to test the correctness of the noise
covariance matrix.  This is performed under the 
requirement that the resultant covariance matrix remain positive
definite; if the matrix becomes non-positive definite for excessively
negative power that power is  given a likelihood of zero as any
negative power is nonphysical.

\section{Results}
\begin{figure}[!t]
\begin{center}
\includegraphics{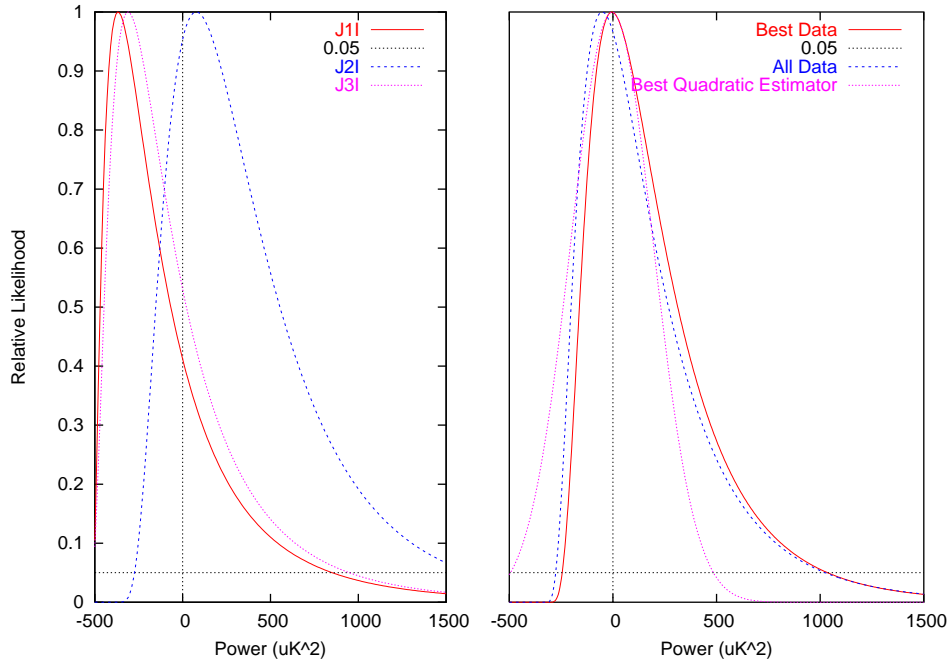}
\caption[Likelihood Curves]{Likelihood curves in $\mu K^2$ for the
E-mode flat band power for each frequency channel in the ``BEST'' data
and for all frequencies combined for the ``BEST'' and ``ALL'' data.
The quadratic estimator for the ``BEST'' data is included as well.}
\label{fig:result}
\end{center}
\end{figure}

The likelihood described above is computed 
for each sub-season, for the union of all sub-seasons where a slope was
removed, and for the union of all data.  Similarly our analysis is
performed for each frequency channel independently and for the union
of all channels.  All results are reported in Table \ref{tab:res}.
Further, likelihood curves of the combined subseasons and channels are
given in Figure \ref{fig:result}. Five points of the integrated
likelihood curve corresponding to conventional definitions for 1 and 2
standard deviations (68\% and 95\% confidence limits) are provided as
well as the band power at which the likelihood curve obtains its maximum.  
This allows for a simple comparison of the various non-Gaussian curves
without requiring a series of figures.

\begin{table} [!ht]
\begin{center}
\small
\begin{tabular}{|c|c|c|c|c|c|c|} \hline
Seas & Chan& 2.5\% &  16\%& max  & 84\%& 95\% \\ \hline
BEST & J1I & -4.600e+02 & -3.800e+02 & -3.700e+02 & 3.000e+02 &
1.290e+03 \\ \hline
BEST & J2I & -1.800e+02 & -2.000e+01 & 8.000e+01 & 8.400e+02 &
1.770e+03 \\ \hline
BEST & J3I & -4.500e+02 & -3.500e+02 & -3.100e+02 & 3.400e+02 &
1.290e+03 \\ \hline
BEST & J1O & -3.200e+02 & -1.600e+02 & -7.000e+01 & 7.600e+02 &
1.790e+03 \\ \hline
BEST & J2O & -2.800e+02 & -1.300e+02 & -6.000e+01 & 8.400e+02 &
2.010e+03 \\ \hline
BEST & J3O & -1.700e+02 & 4.000e+01 & 1.700e+02 & 1.210e+03 &
2.440e+03 \\ \hline
BEST & ALL & -1.800e+02 & -7.200e+01 & -4.000e+00 & 5.000e+02 &
1.128e+03 \\ \hline
ALL & J1I & -4.600e+02 & -3.600e+02 & -3.400e+02 & 4.300e+02 &
1.530e+03 \\ \hline
ALL & J2I & -1.100e+02 & 9.000e+01 & 2.200e+02 & 1.130e+03 &  
2.240e+03 \\ \hline
ALL & J3I & -4.200e+02 & -2.900e+02 & -2.500e+02 & 6.100e+02 &
1.860e+03 \\ \hline
ALL & J1O & -3.000e+02 & -1.300e+02 & -2.000e+01 & 8.500e+02 &
1.910e+03 \\ \hline
ALL & J2O & -2.900e+02 & -1.300e+02 & -8.000e+01 & 9.600e+02 &
2.370e+03 \\ \hline
ALL & J3O & -1.600e+02 & 7.000e+01 & 2.000e+02 & 1.360e+03 & 2.750e+03
\\ \hline
ALL & ALL & -2.120e+02 & -1.120e+02 & -5.200e+01 & 4.760e+02 &
1.148e+03 \\ \hline

\end{tabular}
\caption[Results]{\small \CO results grouped by sub-season and channel.
``seas'' indicates sub-season: first letter is the scan size ({\bf B}ig or
{\bf S}mall) and the second is the ground screen configuration ({\bf
I}nner or {\bf O}uter {\bf G}round {\bf S}creens), BEST is the union of
all data represented in the sub-seasons with only a slope and DC
removed. ``chan'' indicates 
the channel.  The remaining columns are the values in $\mu K^{2}$ for
the indicated values of the integrated likelihood curve.  ``max'' is
the peak location of the curve.}
\label{tab:res}
\end{center}
\end{table}
				   
In order to explore whether or not the data were contaminated by
non-cosmological signals a number of difference tests were performed.
These tests are referred to as jacknife tests or difference map
tests, and though not necessarily optimal they have the benefit of
being both easy to implement and often easy to interpret.

\begin{table}
\begin{center}
\begin{tabular}{|c|c|c|} \hline
\multicolumn{3}{|c|}{Jacknife Tests}  \\ \hline
Cut used        &  Maps generated       & Effects tested for  \\\hline  
Left-Right      & west going/east going & short time scale   \\
                & Azimuth scans         & and scan effects\\\hline 
First-Last      &  first half/last half &long time scale  \\
                &  of each sub-season   & (e.g. thermal, etc.)\\ \hline
Sub Seasonal    & two sub seasons       & sidelobe contamination \& \\
                &                       & long time scale \\ \hline
\multicolumn{3}{|c|}{Rebinning Tests}  \\ \hline
Ra-Az           & Local Sidereal Time and Az    
                & diurnal effects       \\ \hline
Azimuth         & only Azimuth
                & scan synchronous      \\ \hline
\end{tabular}
\caption[Test Description]{This table describes briefly the different
jacknife and binning tests that were performed and states what
effect they are primarily attempting to discriminate.}
\label{tab:jack_desc}
\end{center}
\end{table}

Given a data vector $d_i$ and a noise covariance matrix $N_i$
for each of the $i=0,1$ maps we can form the difference map and noise
covariance 
\begin{eqnarray}
N_d &=&\left[ N_0 + N_1\right], \text{and}  \\
d_d &=& \left[ d_0 - d_1\right].
\end{eqnarray}
The matrices $N_0, N_1$ are actually the generalized noise covariance
matrices and thus include information about constraints.  Therefore if
a constraint is projected out of either map it is projected out of the
difference map as well.  Often the convention of dividing both $N_d$
and $d_d$ by 2 is implemented.  Had we adopted that convention all of
our difference results would be divided by 4 for they are reported in
units of $\mu K^2$.

The tests considered are explained in Table \ref{tab:jack_desc} and
summarized in Table \ref{tab:jacks_performed}.  66 of the indicated 68
tests were passed which is consistent with there being no spurious
signals in the data at the level of several tens of $\mu$K.

\begin{table}
\begin{center}
\begin{tabular}{|c|c|c|c|} \hline
\multicolumn{4}{|c|}{Jacknife Tests}  \\ \hline \hline
Cut used  &  Channels implemented  & Sub-seasons Used    & Total
number 
\\ \hline
Day-Night       & JI, JO        & all   & 2 \\ \hline
Left-Right      & J?I, J?O      & each  & 24 \\ \hline
First-Last      & J?I, J?O      & each  & 24 \\ \hline
Sub Seasonal    & J?I, J?O      & SIGS, SOGS, BIGS  & 18  \\ \hline
\hline \hline
Total           &               &       & 68 \\ \hline
\end{tabular}
\caption[Jacknife Summary] {Summary of the various
combinations of channels and sub-seasons used in each jackknife
test.  ? means vary over each of the 3 channels.  For sub-seasonal
tests only those sub-seasons used in the final analysis (see section 2)
were used.}
\label{tab:jacks_performed}
\end{center}
\end{table}

\section{Conclusion}

After one season of observations totaling approximate 150 hours of
useful data \CO set a 95\% confidence upper limit on polarized celestial
signal of  33.5 $\mu K$ (21.8 $\mu$K 68\% limit) in the $l$-range
$[200-500]$.  It is worth noting that a quadratic estimator approach
to this analysis would have provided a limit of 22.4 $\mu K$, nearly a
factor of two lower (in units of $\mu K^2$).  This emphasizes the need
for explicit calculation of a likelihood curve because the quadratic
estimator is a biased estimator in low signal-to-noise situations.

It is also noteworthy that this limit is the most stringent constraint
on polarized foregrounds in the NCP region on these angular scales.
Because we are observing at a frequency where synchrotron emission is
expected to be the dominant foreground the absence of a signal at this
level bodes well for future CMB polarization experiments.

\section{Acknowledgments}
We thank John Carlstrom for loaning us the HEMT amplifiers used in
this experiment. This research was supported by NSF grants AST-9802851
and AST-9813920 and NASA grant NAG5-11098.  Brian Keating acknowledges
support from the National Science Foundation's Astronomy \&
Astrophysics Postdoctoral Fellowship Program.  Alan Peel acknowledges
support from a NASA/GSRP fellowship.

\bibliographystyle{apj} 
\bibliography{thesis}
\end{document}